\documentclass[ amsmath,amssymb, aps, pra,]{revtex4-2}

\usepackage{graphicx}
\usepackage{dcolumn}
\usepackage{bm}
\usepackage[utf8]{inputenc}      
\usepackage[T1]{fontenc}         
\usepackage[english]{babel}       

\begin{document}

\preprint{APS/123-QED}

\title{4-Pixel NbN Hot-Electron Bolometer Integrated in a Si$_3$N$_4$ Planar Optical Waveguide with On-Chip Fiber-Alignment Trench}

\author{N.A. Vovk}
\affiliation{%
 Institute of Nanotechnology of Microelectronics, Russian Academy of Sciences (INME RAS), 32A Leninsky Prospekt, Moscow 119991, Russia}%

\author{G.A. Matveev	}
\affiliation{%
 Institute of Nanotechnology of Microelectronics, Russian Academy of Sciences (INME RAS), 32A Leninsky Prospekt, Moscow 119991, Russia}%
\author{M.A. Mumlyakov	}
\affiliation{%
 Institute of Nanotechnology of Microelectronics, Russian Academy of Sciences (INME RAS), 32A Leninsky Prospekt, Moscow 119991, Russia}%
\author{M.V. Shibalov	}
\affiliation{%
 Institute of Nanotechnology of Microelectronics, Russian Academy of Sciences (INME RAS), 32A Leninsky Prospekt, Moscow 119991, Russia}%
\author{I.A. Filippov	}
\affiliation{%
 Institute of Nanotechnology of Microelectronics, Russian Academy of Sciences (INME RAS), 32A Leninsky Prospekt, Moscow 119991, Russia}%
\author{I.D. Burkov	}
\affiliation{%
 Institute of Nanotechnology of Microelectronics, Russian Academy of Sciences (INME RAS), 32A Leninsky Prospekt, Moscow 119991, Russia}%
\author{S.D. Perov	}
\affiliation{%
 Institute of Nanotechnology of Microelectronics, Russian Academy of Sciences (INME RAS), 32A Leninsky Prospekt, Moscow 119991, Russia}%
\author{N.V. Porohov	}
\affiliation{%
 Institute of Nanotechnology of Microelectronics, Russian Academy of Sciences (INME RAS), 32A Leninsky Prospekt, Moscow 119991, Russia}%
\author{N.N. Osipov	}
\affiliation{%
 Institute of Nanotechnology of Microelectronics, Russian Academy of Sciences (INME RAS), 32A Leninsky Prospekt, Moscow 119991, Russia}%
\author{M.A. Tarkhov	}
\affiliation{%
 Institute of Nanotechnology of Microelectronics, Russian Academy of Sciences (INME RAS), 32A Leninsky Prospekt, Moscow 119991, Russia}%

\begin{abstract}
In this work, we design and characterize a 4-pixel superconducting hot-electron bolometer (HEB) based on niobium nitride (NbN), integrated with individual planar silicon nitride (Si$_3$N$_4$) waveguides. The implemented architecture enables simultaneous detection of an optical signal in four independent channels. To efficiently couple optical radiation under cryogenic conditions, we employ an edge (end-fire) coupling approach using dedicated U-shaped grooves that provide accurate and stable positioning of an optical fiber with respect to the on-chip waveguide facet. The device responsivity is measured as a function of the HEB operating point. The measured voltage responsivity reaches $3800~\mathrm{V/W}$ at a modulation frequency of $3~\mathrm{GHz}$. We demonstrate detection of optically modulated signals in the gigahertz range. The developed fabrication route is promising for compact integrated receiver systems and low-noise cryogenic microwave transducers, including superconducting nanowire single-photon detectors (SNSPDs).

\end{abstract}

\maketitle


\section{\label{sec:level1}Introduction}

Superconducting niobium nitride (NbN) hot-electron bolometers (HEBs) remain key elements of terahertz heterodyne receivers due to a favorable combination of a relatively high critical temperature, a short electron--phonon relaxation time, and technological compatibility with planar antennas. Together, these factors facilitate efficient coupling and impedance matching of the frontend and, consequently, enable low receiver noise temperatures ($T_\mathrm{rec}$)\cite{Zhang2022,Risacher2018,Khosropanah2007}. The use of nanobridges only a few nanometers thick is a crucial prerequisite for fast electron cooling and efficient mixing up to the upper part of the terahertz band\cite{Khosropanah2007,Puetz2016,Mirzaei2023}. The ultimate performance of HEB mixers is governed by the balance of thermal pathways, the microbridge geometry, and impedance matching to the antenna and the intermediate-frequency (IF) chain; the IF bandwidth $\Delta f_\mathrm{IF}$ is limited by electron--phonon relaxation and electron diffusion\cite{Puetz2016,Puetz2012}. Ongoing progress relies on optimizing the composition and thickness of NbN films, improving substrates and thermal interfaces, and using low-power pumping, which jointly reduce $T_\mathrm{rec}$ and expand $\Delta f_\mathrm{IF}$\cite{Risacher2016,Heyminck2012,Silva2024}.

A major next step in receiver development has been the transition from classical free-space optics to photonic integrated circuits (PICs), where coupling, routing, and processing of radiation are implemented directly on chip\cite{Thomson2016Roadmap,Lin2025OptStructPD}. In conventional optical configurations, the electromagnetic field is delivered to the sensitive element via external optics and discrete matching components, which increases loss, size and alignment sensitivity\cite{Zhao2011FreeSpace,Calkins2013FreeSpace,Meledin2011IEEE}. The maturation of integrated photonics has enabled planar waveguides on Si, SiN, and InP platforms, supporting guided modes and monolithic integration of passive and active elements within a single planar architecture using CMOS-compatible processes\cite{Thomson2016Roadmap}. In such structures, optical radiation is typically coupled into the photonic circuit through grating couplers or edge couplers, which match the fiber mode to the guided waveguide mode\cite{Taillaert2004Grating,Chen2010TiltedFacet}. After coupling, the guided mode propagates along the waveguide, while interaction with the sensitive element is provided by the evanescent field extending into the superconducting or semiconductor region\cite{SnyderLove1983,Haus1984,Schuck2018Nanophotonics}. This mechanism enables stable and reproducible mode coupling over tens of micrometers and reduces alignment sensitivity compared with free-space systems.

Waveguide platforms initially focused on semiconductor photodetectors, where the active region is placed above or near an optical waveguide and absorbs a portion of the guided light through the evanescent tail. Representative examples include Ge $p$--$i$--$n$ photodiodes integrated with silicon waveguides, offering high bandwidth and practical responsivity in the telecom wavelength range\cite{Chen2007GePD}, as well as InGaAs $p$--$i$--$n$ photodetectors heterogeneously integrated with SOI (Silicon on Insulator) waveguides and explicitly using evanescent coupling to deliver optical power into the absorber\cite{RoelkensInGaAsSOI}. Waveguide-coupled III--V compound $p$--$i$--$n$ photodiodes based on InP/InGaAs heterostructures have also been demonstrated, designed for high-speed optical reception and integrated with silicon guiding structures\cite{Pregnolato2022NatComm}. After semiconductor waveguide-integrated photodetectors were established, the same concept was extended to superconducting devices: superconducting nanowire single-photon detectors (SNSPDs) were integrated with NbN\cite{Najafi2015NatCommun}, NbTiN\cite{Schuck2013SciRep}, WSi\cite{McDonald2019APL}, and MoSi\cite{Li2016OE} waveguides, where a nanofilm is placed directly above the guiding structure and absorbs guided light via the evanescent field. These approaches include NbTiN detectors on Si$_3$N$_4$ photonic chips, NbN and WSi SNSPDs in hybrid integrated circuits on silicon and III--V platforms, and MoSi nanowires integrated with planar waveguides and characterized via nano-optical response mapping\cite{Najafi2015NatCommun,Schuck2013SciRep,McDonald2019APL,Li2016OE}. NbTiN detectors on Si$_3$N$_4$ photonic circuits as well as NbN SNSPDs on SOI platforms with integrated grating couplers and low-loss matching structures have been realized using fully CMOS-compatible technologies\cite{Zhao2021SiAWG}.

The next stage in HEB development is the integration of superconducting bolometers with optical waveguide structures, enabling compact chip-level receiver systems for the terahertz and infrared bands. In Mattioli \textit{et al.}\cite{Mattioli2020Nano} and subsequent work by Martini \textit{et al.}\cite{Martini2021OE}, NbN layers were integrated with planar photonic waveguides, where the HEB acts as a linear power detector, in contrast to single-photon SNSPDs. Such architectures demonstrate efficient absorption and conversion of guided radiation into an electrical signal with high speed and sensitivity, confirming the compatibility of NbN HEBs with integrated quantum photonics\cite{Martini2021OE,Torrioli2023OE,Tong2001ISSTT}. Development of this direction includes hybrid and alternative platforms---from MgB$_2$ and NbTiN\cite{Gan2021APL,Karasik2014} to NbN/graphene structures\cite{Gosciniak2024APR}---as well as integrated bolometric detectors based on transparent conducting oxides\cite{Shim2025LSA}. Overall, these studies reflect a broader trend toward waveguide-integrated bolometric systems that combine superconducting electronics with photonic technologies for quantum communications, spectroscopy, and astronomy\cite{Martini2021OE,Torrioli2023OE,Tong2001ISSTT,Shim2025LSA}.

Compared with SNSPDs, NbN-based HEBs offer several fundamental advantages. In particular, the detection mechanism in an HEB is not directly tied to the superconducting energy gap or Cooper-pair breaking. Absorption of radiation heats the electronic subsystem and drives a nonequilibrium electron distribution, making the HEB response substantially less dependent on the energy of an individual photon\cite{Semenov2002}. The speed of an HEB is governed by the cooling dynamics of the electronic subsystem. After rapid electron--electron thermalization, the electron temperature relaxes via electron--phonon coupling, and the characteristic response time is determined by the electron--phonon relaxation rate and, depending on the design, by the phonon escape time into the substrate\cite{Semenov2002,Klapwijk2017,Karasik2014}. In the simplest approximation, the response time can be estimated as:
\[
\tau_{\text{HEB}} \approx \frac{C_e}{G_{e\text{-}ph}},
\]
where $C_e$ is the electronic heat capacity and $G_{e\text{-}ph}$ is the electron--phonon thermal conductance\cite{Zhang2022HEB}. For NbN HEBs, this yields response times on the order of tens of picoseconds, corresponding to gigahertz bandwidths\cite{Klapwijk2017}. In contrast, the speed of an SNSPD is governed by hotspot formation and relaxation and is typically limited to nanosecond time scales\cite{Natarajan2012}.

In many applied scenarios, single-photon detection is not the primary requirement; instead, the key capability is faithful reproduction of high-frequency amplitude modulation of the incident radiation. A representative example is terahertz nondestructive evaluation, including spectroscopy and amplitude-modulated measurements, where bolometric detectors measure the power of quasi-continuous-wave radiation. In such schemes, the decisive parameters are response speed, minimum detectable power, and dynamic range, which determine performance in materials diagnostics, biomedical imaging, and industrial quality control\cite{Jepsen2011LPR,Tonouchi2007NatPhoton,Ferguson2002NatMat}.

Here we demonstrate the integration of a four-pixel array of NbN superconducting HEBs with individual planar Si$_3$N$_4$ waveguides. Each bolometer is optically and electrically isolated and coupled to a dedicated waveguide channel, enabling simultaneous and independent optical readout of multiple channels on a single chip. This architecture represents an important step toward scalable multichannel superconducting receivers for integrated photonics and cryogenic applications. Optical coupling into the waveguides was implemented via an edge-coupling approach enabled by specially formed U-shaped grooves for precise positioning and fixation of the optical fibers on the chip. This coupling scheme provides a broad optical bandwidth, which is critical for cryogenic measurements. Accounting for RF and optical losses, the measured voltage responsivity reaches $3800~\mathrm{V/W}$ at $3~\mathrm{GHz}$. The presence of a response in the gigahertz range indicates potential for high-speed operation; however, the present bandwidth assessment is limited by losses in the RF measurement chain and by the chosen HEB topology. The developed fabrication route is flexible and can be adapted to waveguide-integrated SNSPDs.

Finally, we note that, when placed on a cryogenic stage, the bolometer can serve not only as a detector but also as an active element, acting as a modulator or a signal source in the gigahertz range. This functionality can be achieved by modulating its resistance with high optical power, opening a path toward fully integrated transceiver modules.

\section{\label{sec:level1}HEB Fabrication}

A 4-inch $p^{+}$-type silicon wafer was used as the starting substrate. First, the silicon surface was thermally oxidized to form an SiO$_2$ layer with a thickness of approximately $4~\mu\mathrm{m}$, which served as the lower buffer (bottom cladding) for subsequent optical waveguide (OWG) formation.

A $400~\mathrm{nm}$ Si$_3$N$_4$ layer was deposited on the oxide by plasma-enhanced chemical vapor deposition (PECVD), providing a low-loss planar waveguide platform widely used in integrated photonics. The PECVD Si$_3$N$_4$ waveguide technology was investigated in detail by Dmitriev \textit{et al.}\cite{DmitrievHighPerformanceScNx} and was shown to exhibit high optical stability and low loss.

The overall layout of the integrated superconducting bolometer placed along the optical waveguide is shown in Fig.~1(a). The waveguide geometry was defined by optical lithography followed by plasma etching, yielding planar waveguides with a width of $1500~\mathrm{nm}$ that support guided propagation along the substrate surface with a predominantly quasi-TE mode.

The formation of the waveguide stack with top cladding and bottom buffer layers followed a process similar to that described by Mumlyakov \textit{et al.}\cite{Mumlyakov2024VoidFree}. In particular, the top SiO$_2$ cladding was deposited by PECVD and subsequently chemical--mechanical polished (CMP) to a residual thickness of approximately $140~\mathrm{nm}$ (Fig.~1(b)). This approach yields a smooth, uniform surface free of porosity and defects, which is critical for minimizing optical losses in integrated waveguide structures.

Prior to NbN deposition, the surface was ion-activated by argon ions with energies of several keV in order to clean and modify the substrate surface. This step was performed within the same vacuum cycle as the subsequent magnetron sputtering and was required to improve the NbN--Al interface quality and to ensure a low-resistance ohmic contact. Next, a thin superconducting niobium nitride (NbN) film with a thickness of $9~\mathrm{nm}$ was deposited by magnetron sputtering. Optical lithography and reactive ion etching (RIE) were used to define the topology of the future sensing elements. Aluminum (Al) contact pads with a thickness of $200~\mathrm{nm}$ were fabricated on top by magnetron deposition using a lift-off process.

The Al layout followed the NbN pattern except for the central region, where an uncovered gap was left. As a result, the NbN section not covered by Al formed the active bolometer area as a strip with a width of $1~\mu\mathrm{m}$ and a length of $7~\mu\mathrm{m}$ aligned with the waveguide (see Fig.~1). The final fabrication step was the formation of U-shaped grooves with a depth of approximately $65~\mu\mathrm{m}$ in the silicon substrate using a Bosch deep reactive ion etching process. This depth was chosen to equal one half of the diameter of a standard single-mode fiber (SMF). The grooves enable accurate positioning and fixation of optical fibers for end-fire coupling into the waveguide endface. The waveguide endface was formed at an angle of $8^{\circ}$ relative to the surface normal, which reduces back-reflection at the fiber--waveguide interface. This configuration represents the first experimental demonstration of a grating-free coupling scheme tailored for such cryogenic applications.
\begin{figure*}[!htbp]
    \centering
    \includegraphics[width=0.95\textwidth]{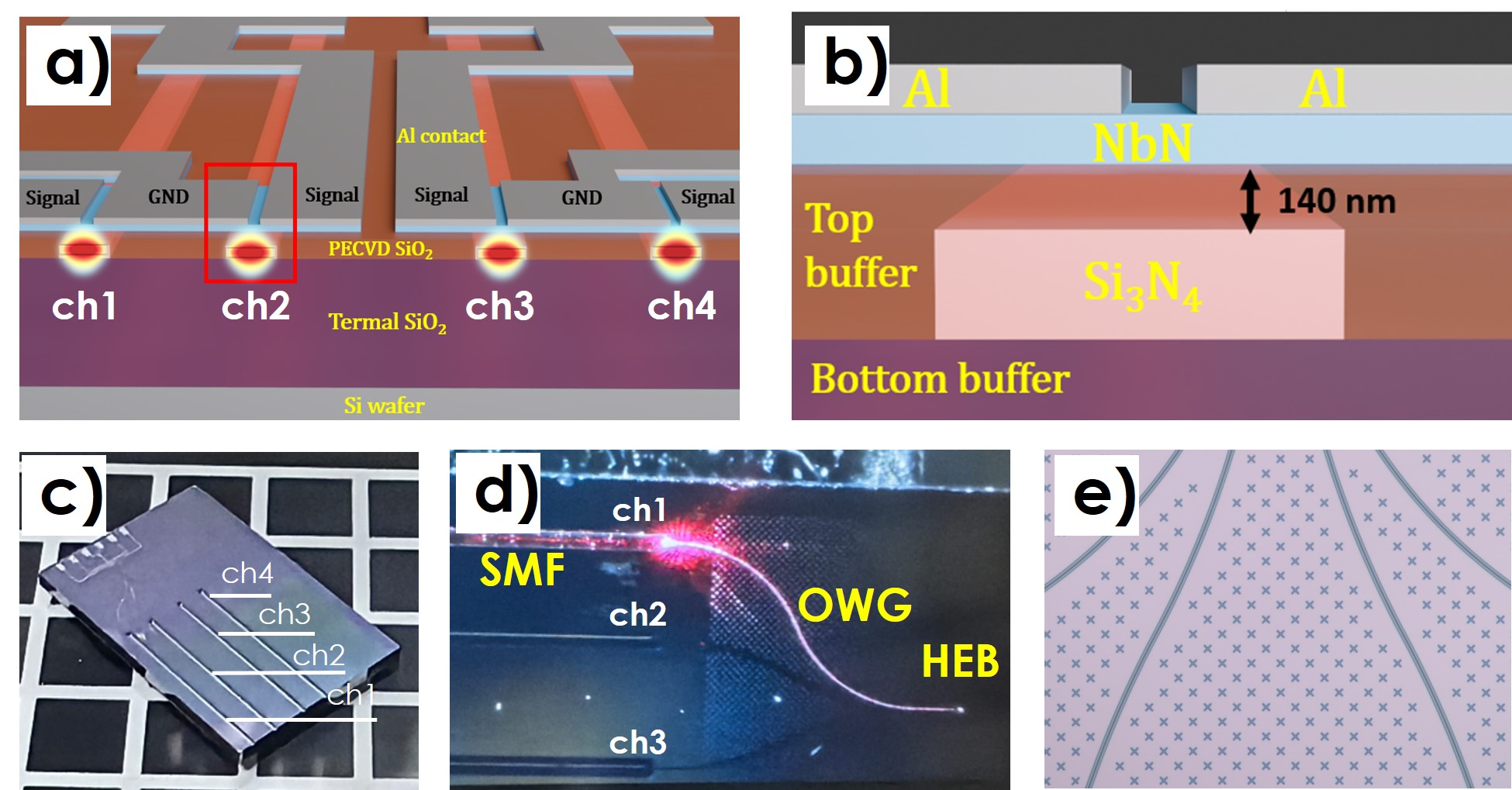}
    \caption{
        (a) Schematic of HEB structures integrated with planar Si$_3$N$_4$ optical waveguides.
        (b) Schematic of the chip endface with the HEB in the waveguide region is outlined in red (a).
        (c) Optical micrograph of the fabricated chip with U-shaped grooves for end-fire coupling.
        (d) SMF-to-planar-waveguide coupling through an on-chip groove.
        (e) Optical micrograph of a waveguide array with cross-shaped optical scatterers.
    }
    \label{fig:HEB_overview}
\end{figure*}

The fabricated chip [Fig.~1(c)] had dimensions of $5.5 \times 8.8~\mathrm{mm}^2$ and contained four independent optical inputs and four integrated bolometers positioned along the corresponding waveguides. This architecture enables independent channel operation with no measurable crosstalk and high optical coupling efficiency. The SMF--OWG alignment quality was verified by launching visible red light into the SMF and observing scattering along the on-chip waveguide, confirming light injection into the planar waveguide [Fig.~1(d)].

In addition, optical scatterers in the form of an array of Si$_3$N$_4$ crosses were patterned near the waveguides to suppress optical crosstalk between adjacent channels [Fig.~1(e)]. These structures provide spatial isolation for waveguides separated by $10~\mu\mathrm{m}$ and mitigate parasitic reflections and mutual coupling.

Finally, we note that the described process for integrating an NbN superconducting layer on a Si$_3$N$_4$ planar waveguide platform is readily adaptable and can be directly adapted to fabricate waveguide-integrated SNSPDs.

\section{\label{sec:level1}Methodology for Measuring the HEB Responsivity}
Since the HEB operates at temperatures below the critical temperature of the NbN superconducting film, electro-optical characterization requires fiber-optic interconnects that remain functional at cryogenic temperatures. During assembly, the optical fiber was aligned and fixed in the on-chip groove (see Fig.~1(d)). The device was then mounted in a closed-cycle Gifford--McMahon cryostat, which cooled the sample to an operating temperature of $2.5~\mathrm{K}$. To reduce intrinsic temperature fluctuations of the HEB detector and suppress external thermal fluctuations during cryogenic measurements, a Teflon spacer was used to partially thermally decouple the sample from the cryostat cold stage.

Figure~2 shows the complete experimental setup for electro-optical characterization of the integrated HEB. Radiation from a continuous-wave telecom laser ($\lambda = 1550~\mathrm{nm}$) passed through a fiber polarizer to ensure efficient coupling to the quasi-TE mode of the planar waveguide and was injected into the PIC via end-fire coupling. Optical intensity modulation was performed using a fiber-coupled electro-optic amplitude modulator based on a LiNbO$_3$ Mach--Zehnder interferometer driven by a Rohde~\&~Schwarz SMW200A RF generator, providing sinusoidal modulation over the target frequency range. A Rof~Electro-Optics fiber modulator (AM series, $\lambda = 1550~\mathrm{nm}$) was used in the experiments.

A Mini-Circuits ZFBT-6GW+ bias-tee was used to apply the DC bias and extract the RF signal, separating the DC and RF components up to $6~\mathrm{GHz}$. The RF output of the bias-tee was connected to a cascade of Mini-Circuits ZX60-3018G-S+ low-noise microwave amplifiers, providing an overall gain in a bandwidth up to $18~\mathrm{GHz}$. The cryostat temperature was stabilized with a precision heater to ensure reproducible measurement conditions. During initial calibration, the HEB operating point was coarsely adjusted by varying the temperature and the bias voltage $U_\mathrm{bias}$ near the optimum. At this stage, the RF signal from the bolometer was fed directly to a Rigol DS70504 digital oscilloscope (5~GHz bandwidth) for preliminary waveform analysis and to verify that the HEB response matched the optical modulation frequency.
\begin{figure}[!htbp]
    \centering
    \includegraphics[width=0.95\linewidth]{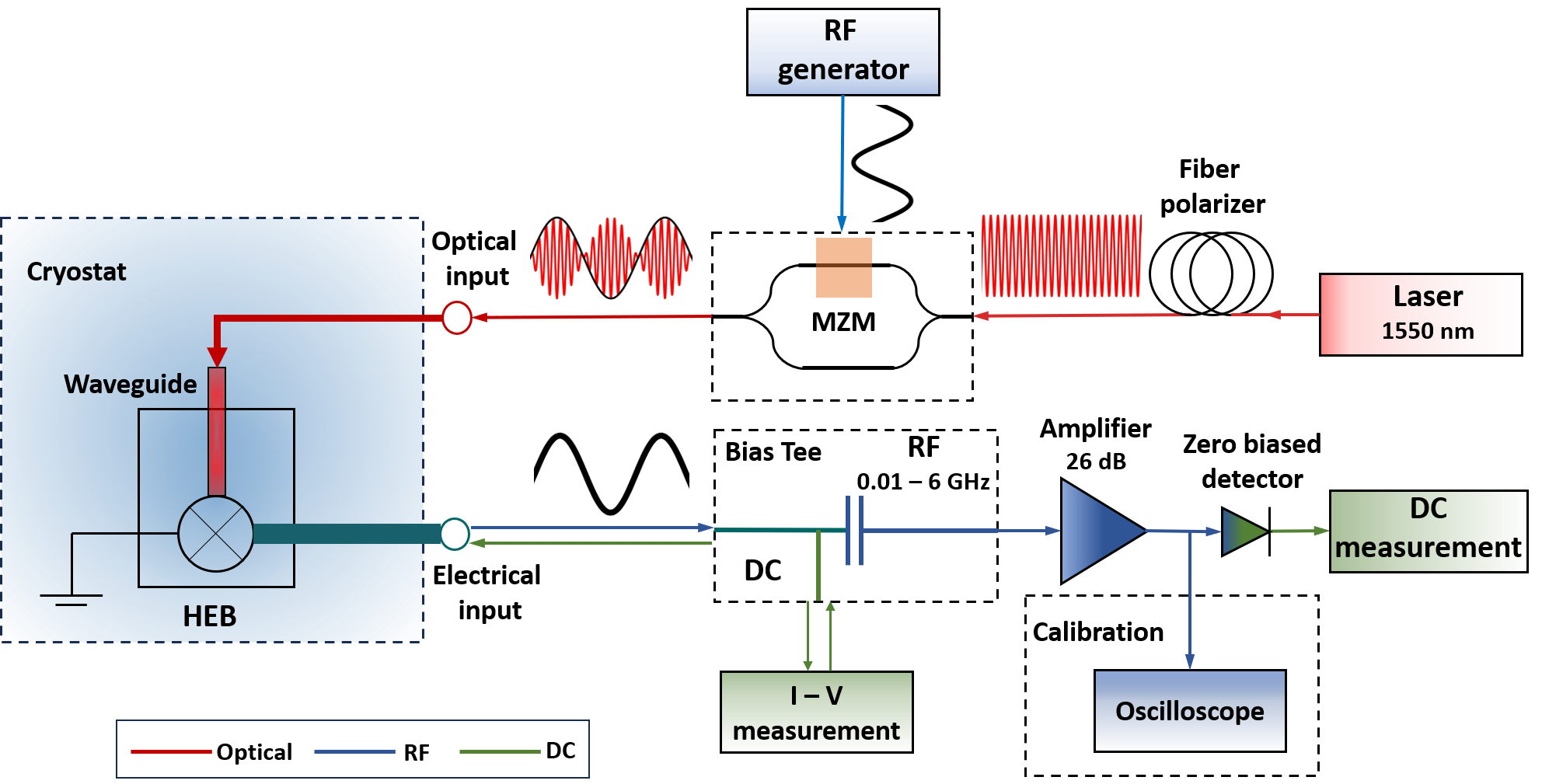}
    \caption{Schematic of the electrical and optical interconnects to the sample inside the cryostat.}
    \label{fig:setup}
\end{figure}
After the signal was identified, quantitative measurements were performed as follows. The RF output of the HEB was connected to a zero-bias Schottky detector (Pasternack PE8011), which rectifies the high-frequency signal into a DC voltage (negative video output). The maximum allowable input power of the detector was $+20~\mathrm{dBm}$, and the operating frequency range was from $10~\mathrm{MHz}$ to $4~\mathrm{GHz}$. The resulting DC voltage was recorded with a Keithley DMM6500 digital multimeter. In parallel, the bias voltage and the current--voltage characteristics were measured using a Keithley~2450 source-measure unit.

As the planar waveguide supports only the TE mode, a polarization dependence was observed: the maximum response exceeded the minimum by approximately $50\%$. All subsequent measurements were carried out with the polarization adjusted for maximum HEB response.

\section{\label{sec:level1}Results}
To evaluate bolometer performance, we measured a series of current--voltage ($I$--$V$) characteristics at different substrate temperatures $T$ under a fixed optical modulation spectrum. Each $I$--$V$ curve was segmented by the bias voltage $U_{\mathrm{bias}}$. For each segment, we determined the amplitude response $V_{\mathrm{HEB}}$, measured by the microwave detector at modulation frequency $f$ and for optical power $P_{\mathrm{optic}}$ propagating in the waveguide. Thus, each portion of the $I$--$V$ family was assigned a value of $V_{\mathrm{HEB}}$ as a relative responsivity at the operating point $(T, U_{\mathrm{bias}})$:

\[
V_{\mathrm{HEB}} = V_{\mathrm{HEB}}(f, T, P_{\mathrm{optic}}, U_{\mathrm{bias}}).
\]

To obtain the system response function $R_{\mathrm{resp}}$ and convert the measured bolometer signal into optical power, we included all losses in the optical and electrical paths. The laser output power at $1550~\mathrm{nm}$ was $P_{\mathrm{laser}} = 40~\mathrm{mW}$. After the optical modulator and polarizer, the power at the cryostat input decreased to $P_{\mathrm{out}} = 0.23~\mathrm{mW}$.

The end-fire coupling efficiency between a single-mode optical fiber ($\lambda = 1550~\mathrm{nm}$, numerical aperture $\mathrm{NA} = 0.14$) and the waveguide was estimated using finite-difference time-domain (FDTD) simulations. The resulting efficiency was $\eta \approx 2\%$. Hence, the optical power reaching the bolometer region is

\[
P_{\mathrm{optic}} = P_{\mathrm{laser}} \cdot \eta = 4.6~\mathrm{\mu W}.
\]

We further estimated the RF loss $L_{\mathrm{RF}} = 0.95~\mathrm{dB}$ in the cryostat lines, measured via the $|S_{21}|$ parameter using a Tektronix TTR506A vector network analyzer while accounting for the connected bias-tee and the amplifier chain in a frequency range up to $5~\mathrm{GHz}$. To convert this loss into a linear voltage attenuation factor $K_{\mathrm{RF}}$, we used

\[
K_{\mathrm{RF}} = 10^{L_{\mathrm{RF}}/20} = 1.115 \quad \text{(at $f = 3~\mathrm{GHz}$)}.
\]

The final expression for the voltage responsivity is

\[
R_{\mathrm{resp}} = \frac{V_{\mathrm{HEB}} \cdot K_{\mathrm{RF}}}{P_{\mathrm{optic}}}.
\]

Figure~3(a) shows the family of $I$--$V$ characteristics at different temperatures $T$, with curve segments color-coded according to $R_{\mathrm{resp}}$ for $P_{\mathrm{optic}} = 4.6~\mathrm{\mu W}$, $f = 3~\mathrm{GHz}$, and the corresponding $(T, U_{\mathrm{bias}})$ values. This representation visualizes how responsivity evolves across the superconducting-to-normal transition and reflects changes in dynamic resistance and thermal relaxation in the transition region. It allows one to identify the bias ranges $U_{\mathrm{bias}}$ where $R_{\mathrm{resp}}$ is maximized.

The black $I$--$V$ curve corresponds to $T = 7.47~\mathrm{K}$ and lies within the region of maximum responsivity.. The pink curve ($T = 7.28~\mathrm{K}$) and the green curve ($T = 8.28~\mathrm{K}$) delineate the boundary of the high-responsivity region. The inset of Fig.~3(a) shows a representative temperature dependence $R_{\mathrm{HEB}}(T)$ with a critical temperature $T_c = 7.5~\mathrm{K}$.
\begin{figure}[!htbp]
    \centering
    \includegraphics[width=\linewidth]{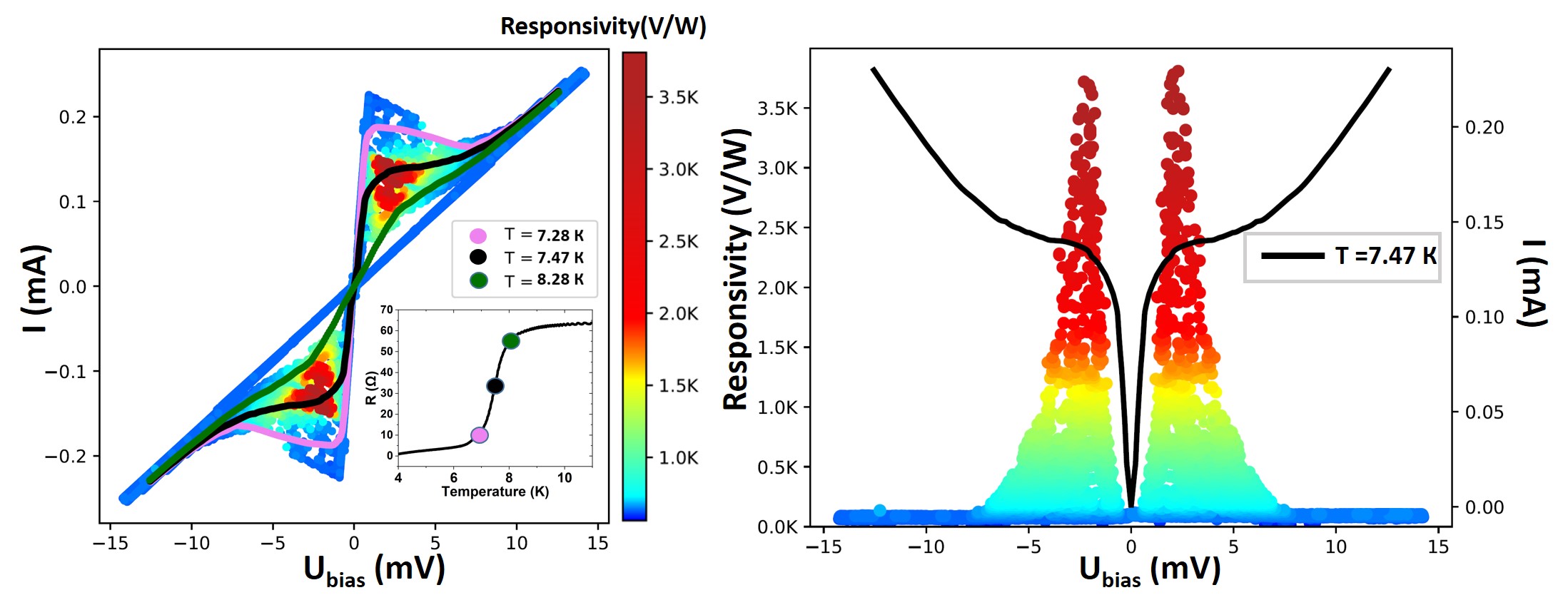}
    \caption{(a) Family of $I$--$V$ characteristics at different temperatures, color-coded according to $R_{\mathrm{resp}}$ for $P_{\mathrm{optic}} = 4.6~\mu\mathrm{W}$, $f = 3~\mathrm{GHz}$, and $U_{\mathrm{bias}}$. The black curve ($T = 7.47~\mathrm{K}$) corresponds to the maximum responsivity; the pink ($T = 7.28~\mathrm{K}$) and green ($T = 8.28~\mathrm{K}$) curves represent boundary regimes. The inset shows $R_{\mathrm{HEB}}(T)$ with $T_c = 7.5~\mathrm{K}$. (b) Family of $R_{\mathrm{resp}}(U_{\mathrm{bias}})$ curves measured at different temperatures $T$, illustrating how the responsivity maximum shifts with temperature. The black line highlights the $I$--$V$ characteristic at $T = 7.47~\mathrm{K}$ corresponding to the maximum-responsivity regime.}
    \label{fig:IV_Rresp}
\end{figure}
Based on these data, we constructed a set of $R_{\mathrm{resp}}(U_{\mathrm{bias}})$ dependencies at different temperatures $T$ [Fig.~3(b)]. These curves show that the position of the responsivity maximum shifts with temperature, enabling identification of the optimal HEB operating point. The maxima of $R_{\mathrm{resp}}$ correspond to regimes where the resistance change per absorbed power is maximized.

The parameters of all four bolometer pixels fabricated on the same chip are summarized in Table~\ref{tab:four_pixels}. All devices exhibit similar critical temperatures and normal-state sheet resistances, and the spread in maximum voltage responsivity does not exceed $\sim 10$--$15\%$, confirming process reproducibility and channel-to-channel consistency.
\begin{table}[!htbp]
\centering
\caption{Parameters of the four bolometer pixels on a single chip}
\label{tab:four_pixels}
\begin{tabular}{c c c c}
\hline
Pixel & $T_c$, K & $R_n$, $\Omega/\square$ & $R_{\mathrm{resp}}^{\max}$, V/W \\
\hline
1 & 7.50 & 63.4 & 3800 \\
2 & 7.47 & 67 & 3500 \\
3 & 7.52 & 65.4 & 3600 \\
4 & 7.45 & 70 & 3400 \\
\hline
\end{tabular}
\end{table}

After complete characterization, we evaluated optical crosstalk caused by radiation propagating from one waveguide into a neighboring bolometer. Despite the small waveguide spacing ($10~\mu\mathrm{m}$), no statistically significant impact on adjacent bolometers was observed, confirming the effectiveness of the implemented optical scatterers.

\section{\label{sec:level1}Conclusions}
In this work, a four-channel waveguide-integrated bolometric detector based on NbN hot-electron bolometers (HEBs) integrated into a planar silicon nitride (SiN) photonic circuit is demonstrated. The waveguide topology enables spatially separated detection of the optical signal by four independent detector pixels. For the first time, optical fibers are assembled directly into the photonic integrated circuit inside the cryostat, providing broadband optical coupling without the use of an interposer or an intermediate chip. To ensure efficient light coupling under cryogenic conditions, an edge-coupling approach with U-shaped grooves is employed, enabling precise and mechanically stable alignment of the optical fiber with the waveguide facet on the chip.

It is shown that an NbN HEB integrated into a planar photonic channel operates both as an RF receiver–mixer and as a waveguide-integrated optical power detector: optical absorption in the active region modifies the electronic state and generates a measurable RF response. The maximum achieved responsivity reaches $3800~\mathrm{V/W}$ at a modulation frequency of $3~\mathrm{GHz}$. The four-channel implementation is promising as a low-noise cryogenic optical modulator of the RF response at the cryostat cold stage for the control of superconducting qubits, with each optical path providing an independent control channel.

\begin{acknowledgments}
The
study was supported by Project No. 125020501540-9 of the
Ministry of Education and Science of the Russian Federation.
Fabrication and technology characterization were carried out
at the large scale facility complex for heterogeneous integration
technologies and silicon+carbon nanotechnologies.
\dots.
\end{acknowledgments}

\bibliography{apssamp}
\end{document}